\documentclass[conference]{IEEEtran}

\newtheorem{theorem}{Theorem}
\newtheorem{lemma}{Lemma}
\newtheorem{corollary}{Corollary}
\newtheorem{definition}{Definition}
\newtheorem{remark}{Remark}
\newtheorem{proposition}{Proposition}
\usepackage{multirow}

\usepackage[cmex10]{amsmath}
\usepackage{amssymb}

\usepackage[dvips]{graphicx}
\usepackage{psfrag}

\IEEEoverridecommandlockouts
\begin{document}

\title{Completion Time in Broadcast Channel and Interference Channel}
\author{\IEEEauthorblockN{Yuanpeng Liu, Elza Erkip}
\IEEEauthorblockA{ECE Department, Polytechnic Institute of NYU, Brooklyn, NY 11201\\
yliu20@students.poly.edu, elza@poly.edu}
\thanks{This work was partially supported by NSF grant No. 0635177.}}
\maketitle

\begin{abstract}
In a multi-user channel, completion time refers to the number of channel uses required for users, each with some given fixed bit pool, to complete the transmission of all their data bits. This paper extends the information theoretic formulation of multi-access completion time to broadcast channel and interference channel, enabling us to obtain the so-called completion time region (CTR), which, analogous to capacity region, characterizes all possible trade-offs between users' completion times. Specifically, for Gaussian broadcast channel (GBC) and Gaussian interference channel (GIC) in the strong/very strong regime, the exact CTR is obtained. For GIC in the weak/mixed regime, an achievable CTR based on the Etkin-Tse-Wang scheme and an outer-bound are obtained.
\end{abstract}

\section{Introduction}
The information theoretic way of approaching a communication network design is usually guided by the assumption that users' data buffers are always full. This assumption greatly simplifies the problem and hence enables a rigorous systematic way to study networks. However, this assumption ignores the bursty nature of real sources and delay considerations, leading to the so-called unconsummated union between information theory and communication networks \cite{Ephremides}.

In \cite{Liu} for multi-access channel (MAC), we considered a periodic source arrival model, where a new block of data of fixed size arrives every $n$ channel uses. Therefore, during each channel block, user's data buffer is not to be replenished by an infinite data reservoir and hence the usual full-buffer assumption is no longer valid. The actual number of channel uses that each user takes to finish its transmission is termed as \emph{completion time}. An information theoretic framework of studying completion time was proposed in \cite{Liu} for MAC. In this paper, we extend the framework to incorporate two important classes of multi-user channels, the broadcast channel (BC) and interference channel (IC) and study the completion time.

Consider the following live video streaming communication scenario as a motivating example. In a multi-user channel, suppose each user wants to either transmit or receive a video stream that is compressed at fixed, but possibly different, rate. Specifically, in the BC setup a common transmitter streams two video sequences to their respective users. In the IC setup, two users stream videos to their respective receivers. The data arrives periodically. In the beginning of each period, there will be a certain number of bits to be received at or transmitted by each user. However due to the casuality constraint, after the completion of the current transmission, new data will not be immediately available until the next period. We model this as the follows: user $i$, $i=1,2$, has $m\tau_i$ bits, with $\tau_i$ corresponding to the compression rate and $m$ corresponding to the number of source samples, to be transmitted in at most $n$ channel uses, where $n$ is assumed to be large enough to allow both transmissions to complete. Let $n_i \leq n$ be the actual number of channel uses that user $i$ spends on communication. The performance metric is \textit{normalized completion time} (referred as completion time hereafter) within a single channel block, which is defined as $n_i/m$ in the limit of large $n_i$ and $m$. Note that in the streaming example, $m$ corresponds to the number of source samples, which is assumed to be the same for both users. In general we can view $m$ as a scaling factor to ensure information theoretic arguments with large block lengths can be invoked. The exact value of $m$ is not important since it will not appear in the characterization of completion time.

The main contributions of this paper are the extension of the information theoretic formulation of completion time, originally proposed in \cite{Liu} for MAC, to BC and IC and, for the Gaussian case, the explicit characterization of the completion time region or inner and outer bounds. Specifically, for GIC in the very strong interference regime, the CTR can be derived directly since GIC reduces into two point-to-point links. For GIC in the strong interference regime, because the capacity region is in the form of Gaussian MAC (GMAC), the derivation of the CTR parallels that in \cite{Liu}. For GIC in the weak and mixed interference regimes, an achievable CTR based on the Etkin-Tse-Wang scheme \cite{Etkin} and an outer-bound are obtained. Toward this end, we adopt the approach used in \cite{Liu}, but generalize the techniques for an arbitrary convex rate region whose boundaries are given by piece-wise linear functions. As for GBC, we adopt a different approach to establish the converse, where the CTR outer-bound is directly obtained by defining a mapping between rate pairs and completion time pairs. We then proceed further to prove the non-convexity of the CTR of GBC by making use of the solution of the weighted sum completion time minimization problem.

Note that in \cite{Rai}, the sum completion time minimization problem for a $K$-user symmetric GMAC was studied. Compared to \cite{Rai}, our result provides a more general formulation for the two-user case, allowing us to consider a variety of utility functions over the CTR, e.g. weighted sum completion time. For GIC, the authors in \cite{Ng} studied the problem of minimizing some convex cost function over the CTR obtained by treating interference as noise, whereas in this paper we adopt an information theoretic approach without restricting decoding strategies to treating interference as noise.

This paper is organized as the follows. In Section II, we define constrained rates for discrete memoryless BC and IC respectively, based on which an information theoretic formulation of completion time is then given. In Section III, we derive the CTR for GBC. In Section IV, we discuss the CTR for GIC case by case. The paper is concluded in Section V.

\emph{Notation}: Denote $\gamma(x)=\frac{1}{2}\textrm{log}_2(1+x)$. Also $X_{k,i}^j=(X_{k,i},...,X_{k,j})$ for $i\leq j$ and $X_k^j=X_{k,1}^j$. $X_{k,i}^j$ does not appear if $i>j$. $[X]^+=\max\{X,0\}$. We use bold font for vectors and calligraphic font for regions.

\section{Problem Formulation}
In this section, we first extend the definition of constrained rates proposed in \cite{Liu} to include broadcast channel and interference channel in Section II.A and II.B respectively. We then define completion time in Section II.C, which is common for both channels.

\subsection{Constrained Rate for Broadcast Channel}
Consider a two-user discrete memoryless broadcast channel (DMBC) $(\mathcal{X},p(y_1,y_2|x),\mathcal{Y}_1\times\mathcal{Y}_2)$ with individual message sets, where $\mathcal{X}$ is the input alphabet, $\mathcal{Y}_1$ and $\mathcal{Y}_2$ are the channel output alphabets and $p(y_1,y_2|x)$ is the channel transition probability. Let $n_i$, $i=1,2$, be the number of channel uses user $i$'s codebook spans. Denote $n=\max\{n_1,n_2\}$, $\pi_1=\arg_{i=1,2}\min\{n_i\}$, $\pi_2=\arg_{i=1,2}\max\{n_i\}$, and $c=n_1/n_2$. We will let $n_1$ and $n_2$ vary with $c$ fixed.

\begin{definition}
A $(M_1,M_2,n,c)$ code consists of message sets: $\mathcal{W}_i=\{1,...,M_i\}$, an encoding function,
\begin{align*}
    X:(\mathcal{W}_1\times\mathcal{W}_2)&\rightarrow \mathcal{X}^{n_{\pi_1}}\\
       \mathcal{W}_{\pi_2}&\rightarrow\mathcal{X}_{n_{\pi_1}+1}^n
\end{align*}
and two decoding functions $g_i:\mathcal{Y}_i^{n_i}\rightarrow \mathcal{W}_i$, $i=1,2$.

Note that the codeword can be viewed as consisting of two parts. The first part $\mathcal{X}^{n_{\pi_1}}$ is determined by the messages of both users while the second part, $\mathcal{X}_{n_{\pi_1}+1}^n$, is solely determined by user $\pi_2$'s message.
\end{definition}

The sender independently chooses an index $W_i$ uniformly from $\mathcal{W}_i$ and sends the corresponding codeword. The average error probability for the $(M_1,M_2,n,c)$ code is
\begin{equation*}
    P_e=\textrm{Pr}(g_1(Y_1^{n_1})\neq W_1 \textrm{ or }g_2(Y_2^{n_2})\neq W_2).
\end{equation*}
\begin{definition}
The \textit{$c$-constrained rates} of $(M_1,M_2,n,c)$ code are, for $i=1,2$,
\begin{equation}
\label{ratedef}
    R_i=\frac{\log_2(M_i)}{n_i} \quad \textrm{bits per channel use}.
\end{equation}
\end{definition}

The $c$-constrained rate pair $(R_1,R_2)$ is said to be \textit{achievable} if there exits a sequence of $(M_1,M_2,n,c)$ codes with $P_e\rightarrow 0$ as $n_1,n_2\rightarrow \infty$ with $c$ fixed. The \textit{c-constrained rate region}, denoted by $\mathcal{R}_c$, is the set of achievable $c$-constrained rate pairs for a given coding scheme. The \textit{c-constrained capacity region} $\mathcal{C}_c$ is the closure of all such $\mathcal{R}_c$.

\begin{remark}
We use the term ``$c$-constrained rate (capacity) region'' to emphasize the fact that user $i$'s effective codeword length is constrained by $n_i$ channel uses over which $R_i$ is defined and the rate (capacity) region is hence a function of $c=n_1/n_2$. Also note that $\mathcal{R}_1$ ($\mathcal{C}_1$) is the standard rate (capacity) region, where $n_1=n_2$. For the rest of this paper, the term ``rate (capacity) region'' refers to standard rate (capacity) region.
\end{remark}

\subsection{Constrained Rate for Interference Channel}
Consider a two-user discrete memoryless interference channel (DMIC) $(\mathcal{X}_1\times\mathcal{X}_2,p(y_1,y_2|x_1,x_2), \mathcal{Y}_1\times \mathcal{Y}_2)$, where $\mathcal{X}_1,\mathcal{X}_2$ are the input alphabets, $\mathcal{Y}_1,\mathcal{Y}_2$ are the channel output alphabets and $p(y_1,y_2|x_1,x_2)$ is the channel transition probability. For $i=1,2$, let $\bar{i}=\{1,2\}\setminus i$ and define
\begin{align*}
    R_i^{\textrm{IC}}=\max_{p_{X_i}} I(X_i;Y_i|X_{\bar{i}}=\phi_{\bar{i}}), \quad i=1,2,
\end{align*}
where $\phi_{\bar{i}}=\arg\max_{\phi\in\mathcal{X}_{\bar{i}}}\max_{p_{X_i}} I(X_i;Y_i|X_{\bar{i}}=\phi)$. One can view $\phi_{\bar{i}}$ as the symbol that ``opens'' up the channel from from transmitter $i$ to receiver $i$ the most.

\begin{definition}
A $(M_1,M_2,n,c)$ code consists of message sets: $\mathcal{W}_i=\{1,...,M_i\}$, two encoding functions,
\begin{align*}
    X_i:\mathcal{W}_i\rightarrow (\mathcal{X}_i^{n_i},\phi_{i,{n_i+1}}^n)\quad \textrm{for } i=1,2
\end{align*}
and two decoding functions,
\begin{align*}
    g_i:\mathcal{Y}_i^{n_i}\rightarrow \mathcal{W}_i\quad \textrm{for } i=1,2.
\end{align*}
Note that user $i$ will send $\phi_i$ during the $n-n_i$ symbols at the end of its codeword.
\end{definition}

The remaining definitions for the error probability and constrained rates follow II.A exactly.

\subsection{The Notion of Completion Time}
Consider either a DMBC or DMIC, where there are $m\tau_i$, $i=1,2$, bits to be received at or transmitted by each user.

\begin{definition}
\label{delaydef}
We define the \textit{normalized completion time} as $d_i=n_i/m$, where $n_i$ is the actual number of channel uses that user $i$ spends on transmitting $m\tau_i$ bits.
\end{definition}

Because of the relation $\log_2(M_i)=n_iR_i=m\tau_i$ in (\ref{ratedef}), where $R_i$ is the $c$-constrained rate, we have $d_i=\tau_i/R_i$. Completion time pair $(d_1,d_2)$ is said to be \textit{achievable} if $(\tau_1/d_1,\tau_2/d_2)$ is an achievable $c$-constrained rate pair, i.e. $(\tau_1/d_1,\tau_2/d_2)\in \mathcal{R}_c$ where $c=n_1/n_2=d_1/d_2$. The achievable completion time region for a given coding scheme is $\mathcal{D}=\{(d_1,d_2)|(\tau_1/d_1,\tau_2/d_2)\in \mathcal{R}_{d_1/d_2}\}$. Analogous to capacity region, we can also define the overall completion time region $\mathcal{D}^*$ as the union of all achievable completion time regions, or equivalently $\mathcal{D}^*=\{(d_1,d_2)|(\tau_1/d_1,\tau_2/d_2)\in \mathcal{C}_{d_1/d_2}\}$. Notice that the definition of $\mathcal{D}^*$ does not involve the convex hull operation as opposed to the capacity region. This is because $\mathcal{D}^*$ may not be convex, as shown in \cite{Liu} and later in this paper Proposition \ref{BCnoconvex}.

\section{Completion Time Region for Gaussian Broadcast Channel}
In this section, we first consider a general discrete memoryless degraded broadcast channel and derive the $c$-constrained capacity region in Section III.A. We then establish the completion time region for GBC in Section III.B. In III.C, we solve the weighted sum completion time minimization problem and prove the non-convexity of the CTR for GBC.

\subsection{Constrained Capacity Region for Degraded BC}
Since stochastic degradedness and physical degradedness are interchangeable for broadcast channel \cite{Cover}, the term ``degraded'' used in this paper implicitly refers to physically degraded, otherwise stated. We first present the $c$-constrained capacity region of degraded BC and then specialize it to the Gaussian case, which belongs to the class of stochastically degraded BC. Lemma \ref{raterelation} and Theorem \ref{CCapaRegionBC}, which will be stated next, reveal the connection between the $c$-constrained rate (capacity) region and the standard one.
\begin{lemma}
\label{raterelation}
The $c$-constrained rate pair $(R_1,R_2)$ is achievable, for some $c$, if:
\begin{enumerate}
\item{$c\leq 1$,}
$R_2$ can be decomposed into $R_2'$ and $R_2''$: $R_2=cR_2'+(1-c)R_2''$, such that $(R_1,R_2')\in\mathcal{C}_1$, $R_2''\leq R_2^{\textrm{BC}}$;

\item{$c\geq 1$,}
$R_1$ can be decomposed into $R_1'$ and $R_1''$: $R_1=\frac{1}{c}R_1'+(1-\frac{1}{c})R_1''$, such that $(R_1',R_2)\in\mathcal{C}_1$, $R_1''\leq R_1^{\textrm{BC}}$,

\end{enumerate}
where $R_i^{\textrm{BC}}$ is defined as
\begin{equation}
    R_i^{\textrm{BC}}=\max_{p_{X}} I(X;Y_i), \quad i=1,2\label{RBC}.
\end{equation}
\end{lemma}
\begin{IEEEproof}
The proof parallels that of \cite[Lemma.1]{Liu}.
\end{IEEEproof}

To avoid confusion, hereafter we use lower-case $r$ and upper-case $R$ to refer to the standard and constrained rates respectively.
\begin{theorem}
\label{CCapaRegionBC}
The $c$-constrained capacity region $\mathcal{C}_c$ for a degraded broadcast channel, where $Y_2$ is degraded w.r.t. $Y_1$, is the set of rate pairs $(R_1,R_2)$ satisfying
\begin{enumerate}
\item{$c\leq 1$, $\left(R_1,\left[\tfrac{1}{c}R_2-(\tfrac{1}{c}-1)R_2^{\textrm{BC}}\right]^+\right)\in\mathcal{C}_1$;}
\item{$c\geq 1$, $\left(\left[cR_1-(c-1)R_1^{\textrm{BC}}\right]^+,R_2\right)\in\mathcal{C}_1$,}
\end{enumerate}
where $\mathcal{C}_1$, the degraded DMBC capacity region, is the set of all $(r_1,r_2)$ pairs satisfying
\begin{align*}
    r_1&\leq I(X;Y_1|U),\\
    r_2&\leq I(U;Y_2),
\end{align*}
for some joint distribution $p(u)p(x|u)(y_1,y_2|x)$, with the auxiliary random variable $U$ cardinality bounded by $|\mathcal{U}|\leq \min\{|\mathcal{X}|,|\mathcal{Y}_1|, |\mathcal{Y}_2|\}$.
\end{theorem}
\begin{IEEEproof}
The proof is relegated to Appendix \ref{ProofCCapaRegionBC}.
\end{IEEEproof}

\begin{remark}
The constrained capacity achieving scheme can be viewed as consisting of two phases. In the first phase when the codeword carries both users' messages, the BC capacity-achieving scheme is employed. In the second phase when the codeword carries only user $\pi_2$'s message, the coding scheme that achieves the point-to-point capacity for user $\pi_2$ is employed.
\end{remark}

Next let us consider a two-user GBC:
\begin{align*}
    Y_1 &= h_1X + Z_1,\\
    Y_2 &= h_2X + Z_2,
\end{align*}
where $Z_i\sim\mathcal{N}(0,1)$, $i=1,2$, is the i.i.d. Gaussian noise process and inputs are subject to a per symbol power constraint: $E[X^2]\leq P$. Without loss of generality, we assume $h_1\geq h_2$. Hence $Y_2$ is stochastically degraded w.r.t. $Y_1$.
\begin{corollary}
\label{coroGBC}
The $c$-constrained capacity region of two-user GBC is the set of non-negative rate pairs $(R_1,R_2)$ satisfying:
\begin{enumerate}
\item{$c\leq 1$, $\left(R_1,\left[\tfrac{1}{c}R_2-(\tfrac{1}{c}-1)R_2^*\right]^+\right)\in\mathcal{C}_1^G$;}
\item{$c\geq 1$, $\left(\left[cR_1-(c-1)R_1^*\right]^+,R_2\right)\in\mathcal{C}_1^G$,}
\end{enumerate}
where $R_i^*=\gamma(h_i^2P)$ and $\mathcal{C}_1^G$, the capacity region of GBC, is the set of non-negative rate pairs satisfying
\begin{align*}
    r_1\leq \gamma(h_1^2P_1),\quad r_2\leq \gamma(h_2^2P) - \gamma(h_2^2P_1),
\end{align*}
where $0\leq P_1\leq P$.
\end{corollary}

\subsection{Completion Time Region}
An achievable completion time pair $\mathbf{d}=(d_1,d_2)$ is defined in terms of $c$-constrained rate pair, which in return depends on $\mathbf{d}$ through $c=d_1/d_2$. Hence it is easy to check for a given $\mathbf{d}$ whether or not it is achievable, but difficult to directly compute all pairs of $\mathbf{d}\in\mathcal{D}^*$ using the definition, because of this recursive dependence. Another difficulty in determining $\mathcal{D}^*$ is that it is not convex for GBC, as we shall show later in Proposition \ref{BCnoconvex}. Therefore we take a different approach. We characterize two sub-regions of $\mathcal{D}^*$ seperately and the union of the two will give us to $\mathcal{D}^*$. In the following, we first show that the sub-regions are always convex.

\begin{proposition}
\label{convex}
$\mathcal{D}^*$ contains two convex sub-regions, $\mathcal{D}_1^*$ and $\mathcal{D}_2^*$, where
\begin{align*}
    \mathcal{D}_{1}^*&=\mathcal{D}^*\bigcap\{(d_1,d_2)|d_1\leq d_2\},\\
    \mathcal{D}_{2}^*&=\mathcal{D}^*\bigcap\{(d_1,d_2)|d_1\geq d_2\}.
\end{align*}
\end{proposition}

\begin{IEEEproof}
See the proof \cite[Proposition.1]{Liu}.
\end{IEEEproof}

Essentially in the proof of Proposition \ref{convex}, we show that for any two given achievable completion time pairs $\mathbf{d}$ and $\mathbf{d}'$, if $\mathbf{d}$ and $\mathbf{d}'$ lie on the same side with respect to the line $d_1=d_2$, then we can always construct a coding scheme such that the new scheme achieves the convex combination of $\mathbf{d}$ and $\mathbf{d}'$.

Theorem \ref{CCapaRegionBC} together with Lemma \ref{raterelation} suggests that any $c$-constrained rate pair $(R_1,R_2)$ can be expressed in terms of standard rate pair. When $c=d_1/d_2\leq 1$, $R_1=r_1$ and $R_2=cr_2+(1-c)R_2''$, where $(r_1,r_2)\in\mathcal{C}_1$ and $R_2''\leq R_2^{\textrm{BC}}$. Substituting $R_i=\tau_i/d_i$ and $c=d_1/d_2$, we have the following relations:
\begin{align}
\label{d1}
    d_1=\tfrac{\tau_1}{r_1},\quad d_2=\tfrac{\tau_2}{R_2''}+\tfrac{(R_2''-r_2)\tau_1}{R_2''r_1},
\end{align}
where $d_1\leq d_2$ reduces to the condition $\frac{r_2}{r_1}\leq \frac{\tau_2}{\tau_1}$. Similarly for $c\geq 1$, we have
\begin{align}
\label{d2}
    d_1=\tfrac{\tau_1}{R_1''}+\tfrac{(R_1''-r_1)\tau_2}{R_1''r_2},\quad d_2=\tfrac{\tau_2}{r_2},
\end{align}
where $R_1''\leq R_1^{\textrm{BC}}$ and $d_1\geq d_2$ reduces to $\frac{r_2}{r_1}\geq \frac{\tau_2}{\tau_1}$. One can think of equations (\ref{d1}) and (\ref{d2}) as functions that map a rate pair to a completion time pair depending on whether $d_1\leq d_2$ or $d_1\geq d_2$. Hence we use $\mathbf{d}_i(\mathbf{r})$ to denote the completion time pair $\mathbf{d}=(d_1,d_2)$ mapped from $\mathbf{r}=(r_1,r_2)$ using (\ref{d1}) if $i=1$ and $\mathbf{r}\in\mathcal{C}_{1,1}^G$, and (\ref{d2}) if $i=2$ and $\mathbf{r}\in\mathcal{C}_{1,2}^G$, where we define $\mathcal{C}_{1,1}^G$ and $\mathcal{C}_{1,2}^G$ as
\begin{align*}
    \mathcal{C}_{1,1}^G&=\mathcal{C}_1^G\bigcap\{(r_1,r_2)|\tfrac{r_2}{r_1}\leq \tfrac{\tau_2}{\tau_1}\},\\
    \mathcal{C}_{1,2}^G&=\mathcal{C}_1^G\bigcap\{(r_1,r_2)|\tfrac{r_2}{r_1}\geq \tfrac{\tau_2}{\tau_1}\}.
\end{align*}

\begin{figure}[htb]
    \centering
    \includegraphics[width=90mm]{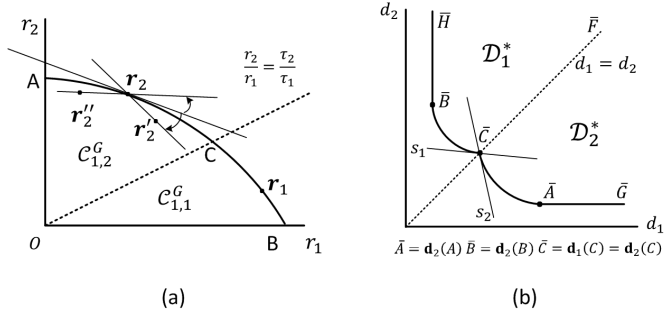}
    \caption{Gaussian broadcast channel: (a) capacity region; (b) completion time region.}
    \label{BCCTRegion}
\end{figure}

Referring to Fig. \ref{BCCTRegion}(a), let $C$ denote the intersection of the line $\frac{r_2}{r_1}=\frac{\tau_2}{\tau_1}$ and the capacity region boundary. This is obtained by substituting $r_1=\gamma(h_1^2P_1)$ and $r_2=\gamma(h_2^2P)-\gamma(h_2^2P_1)$ into the line equation resulting in $P_1=P_1'\in[0,P]$.

\begin{theorem}
\label{CTRGBC}
The completion time region of two-user GBC, depicted in Fig. \ref{BCCTRegion}(b), is given by $\mathcal{D}^*=\mathcal{D}_1^*\bigcup\mathcal{D}_2^*$, where
\begin{align*}
    \mathcal{D}_1^*=\left\{
    \begin{aligned}
        &(d_1,d_2)\in \mathbb{R}^2_+: \textrm{ for } P_1\in[P_1',P]\\
        &d_1\geq \tfrac{\tau_1}{\gamma(h_1^2P_1)},\ d_2\geq d_1,\\
        &d_2\geq \tfrac{\tau_2}{\gamma(h_2^2P)} + \tfrac{\gamma(h_2^2P_1)\tau_1}{\gamma(h_2^2P)\gamma(h_1^2P_1)}
     \end{aligned}
     \right\},
\end{align*}
and
\begin{align*}
    \mathcal{D}_2^*=\left\{
    \begin{aligned}
        &(d_1,d_2)\in \mathbb{R}^2_+: \textrm{ for } P_1\in[0,P_1']\\
        &d_1\geq \tfrac{\tau_1}{\gamma(h_1^2P)} + \tfrac{[\gamma(h_1^2P)-\gamma(h_1^2P_1)]\tau_2} {\gamma(h_1^2P)[\gamma(h_2^2P)-\gamma(h_2^2P_1)]},\\
        &d_2\geq \tfrac{\tau_2}{\gamma(h_2^2P)-\gamma(h_2^2P_1)},\ d_1\geq d_2,
     \end{aligned}
     \right\}.
\end{align*}

\end{theorem}

\begin{IEEEproof}
We first consider $\mathcal{D}_1^*$. For the converse, we consider the mapping defined in (\ref{d1}). Notice that $d_2$ can be alternatively expressed as $d_2=\frac{\tau_2r_1-\tau_1r_2}{R_2''r_1}+\frac{\tau_1}{r_1}$, which is a decreasing function of $R_2''$, since $\frac{r_2}{r_1}\leq \frac{\tau_2}{\tau_1}$. To obtain a lower-bound, we set $R_2''=\gamma(h_2^2P)$. Referring to Fig. \ref{BCCTRegion}(a), any point in $\mathcal{C}_{1,1}^G$ is upper-bounded by some point on the curve $BC$. The curve $\bar{B}\bar{C}$ in Fig. \ref{BCCTRegion}(b) is obtained by mapping every rate point on $BC$ to a completion time point through (\ref{d1}) with $R_2''=\gamma(h_2^2P)$. It's clear that the curve $\bar{B}\bar{C}$ is a lower-bound because $d_1,d_2$ in (\ref{d1}) are decreasing functions of $r_1,r_2$. Together with $d_1\leq d_2$, $\mathcal{D}_1^*$ is a lower-bound.

We now prove the achievability of $\mathcal{D}_1^*$. First, any point on curve $\bar{B}\bar{C}$ is achievable, since curve $\bar{B}\bar{C}$ is obtained by mapping from curve $BC$ in the achievable rate region. Second any point on the ray $\bar{B}\bar{H}$ is achievable. This is because we can use the same codebooks designed for achieving $\bar{B}$ but decrease the rate of user 2 by only using part of the codewords, resulting in the same $d_1$ but a larger $d_2$. For the same reason, any point on the ray $\bar{C}\bar{F}$ is also achievable (here we keep the same codebooks but decrease the rates for both users by the same amount). At last, any inner point can be expressed as the convex combination of two points on the boundary and hence is also achievable due to Proposition \ref{convex}.

Using the same argument, we can prove for $\mathcal{D}_2^*$. Overall $\mathcal{D}^*=\mathcal{D}_1^*\bigcup\mathcal{D}_2^*$.
\end{IEEEproof}

\subsection{Minimum Weighted Sum Completion Time}
Network design often incorporates the goal of optimizing a certain utility function, which, for example, can be a function of users' rates. The completion time region, which characterizes all possible trade-offs between users' completion times, allows one to compute utilities that are functions of users' completion times. In this subsection, we intend to solve the following weighted sum completion time minimization problem:
\begin{align}
    \textrm{minimize    }&\quad d_s=wd_1+\bar{w}d_2 \label{optoverall}\\
    \textrm{subject to  }&\quad (d_1,d_2)\in\mathcal{D}^* \notag
\end{align}
where $\bar{w}=1-w$ and $w\in[0,1]$. As Fig. \ref{BCCTRegion}(b) shows, $\mathcal{D}^*$ is not convex, which we shall prove later in Proposition \ref{BCnoconvex}. Hence it's more convenient to consider problem (\ref{optorigin}), where the feasible sets are convex. Clearly for any given weight, the solution of (\ref{optoverall}) is immediately induced by the minimum of those for (\ref{optorigin}).
\begin{align}
    \textrm{minimize    }&\quad d_s=wd_1+\bar{w}d_2 \label{optorigin}\\
    \textrm{subject to  }&\quad (d_1,d_2)\in\mathcal{D}_{i}^*,\ i=1,2 \notag
\end{align}
Note that $\mathcal{D}_i^*$ given by Theorem \ref{CTRGBC} is expressed in terms of power variable $P_1$. In principle we could first eliminate $P_1$ and alternatively write $d_2$ as a function of $d_1$, then use the convexity of $\mathcal{D}_i^*$ to solve (\ref{optorigin}) by taking the derivative. However in the following, we take a different approach and focus on $\mathcal{C}_1^G$ instead of $\mathcal{D}_i^*$. By using a line rotation argument, we show that every boundary point of $\mathcal{C}_1^G$ uniquely minimizes $d_s$ for a weight that is related to the tangent line of $\mathcal{C}_1^G$ at that point. The reason of taking this indirect approach is that the same geometric argument will also be used to obtain achievable CTR of GIC in the weak and mixed interference regime in Section IV.C.

Let us first transform (\ref{optorigin}) into an equivalent problem using the mapping defined in (\ref{d1}), (\ref{d2}). Define
\begin{align}
\label{Di}
    D_1=\bar{w}\tfrac{\tau_2}{R_2^*}+\tfrac{\tau_1(R_2^*-\bar{w}r_2)}{R_2^*r_1}, \ D_2=w\tfrac{\tau_1}{R_1^*}+\tfrac{\tau_2(R_1^*-wr_1)}{R_1^*r_2},
\end{align}
where $D_i(\mathbf{r})$ denotes $D_i$ evaluated at $\mathbf{r}=(r_1,r_2)$.

\begin{proposition}
\label{eq}
The following optimization problem is equivalent to (\ref{optorigin}):
\begin{align}
    \textrm{minimize    }&\quad D_i(\mathbf{r}) \label{newopt}\\
    \textrm{subject to  }&\quad \mathbf{r}\in\mathcal{C}_{1,i}^G, \ i=1,2\notag
\end{align}
\end{proposition}

\begin{IEEEproof}
Let's first consider (\ref{optorigin}) with $i=1$, i.e. $d_1\leq d_2$. Without loss of generality, consider (\ref{d1}) with $R_2''=R_2^*$, i.e. letting user 2 transmit at the maximum point to point rate in the second phase to minimize its delay. A rate pair $\mathbf{r}$ can be mapped to a completion time pair via (\ref{d1}) resulting in $d_s=D_1(\mathbf{r})$ for $\mathbf{r}\in\mathcal{C}_{1,1}^G$. Similarly for $i=2$, i.e. $d_1\geq d_2$, $\mathbf{r}$ can be mapped to a completion time pair via (\ref{d2}) with $R_1''=R_1^*$ and $d_s=D_2(\mathbf{r})$ for $\mathbf{r}\in\mathcal{C}_{1,2}^G$.
\end{IEEEproof}

\begin{lemma}
\label{tangent}
Denote the tangent line of $\mathcal{C}_1^{G}$ at a boundary point $\mathbf{r}=(r_1,r_2)$ by $ar_1+br_2=1$. Then we have
\begin{align*}
    a=\frac{g}{r_2+gr_1},\quad b=\frac{1}{r_2+gr_1},
\end{align*}
where $r_1=\gamma(h_1^2P_1)$, $r_2=\gamma(h_2^2P)-\gamma(h_2^2P_1)$ and
\begin{align*}
    g=\frac{1/h_1^2+P_1}{1/h_2^2+P_1}.
\end{align*}
Define $w_1(\mathbf{r})=1-bR_2^*,\ w_2(\mathbf{r})= aR_1^*$. Then $w_1(\mathbf{r})\in[0,1)$ and $w_2(\mathbf{r})\in(0,1]$.
\end{lemma}

\begin{IEEEproof}
The proof is relegated to Appendix \ref{Prooftangent}.
\end{IEEEproof}

\begin{theorem}
\label{miniBC}
The solution to the optimization problem (\ref{optorigin}) is summarized in Table \ref{table}, where $\mathbf{r}_1$ ($\mathbf{r}_2$), referring to Fig. \ref{BCCTRegion}(a), is an arbitrary boundary point of $\mathcal{C}_1^G$ that lies between $B$ ($C$) and $C$ ($A$).
\begin{center}
\begin{table}[htbp]
\caption{}
\label{table}
\hfill{}
\begin{tabular}{|c|c|c|c|}

\hline
\multirow{2}{*}{$i=1$} &  $w\in[0,w_1(C)]$   & $w=w_1(\mathbf{r}_1)$         & $w\in[w_1(B),1]$  \\\cline{2-4}
                       &  $\mathbf{d}_1(C)$  & $\mathbf{d}_1(\mathbf{r}_1)$  & $\mathbf{d}_1(B)$   \\\hline
\multirow{2}{*}{$i=2$} &  $w\in[0,w_2(A)]$   & $w=w_2(\mathbf{r}_2)$         & $w\in[w_2(C),1]$  \\\cline{2-4}
                       &  $\mathbf{d}_2(A)$  & $\mathbf{d}_2(\mathbf{r}_2)$  & $\mathbf{d}_2(C)$  \\
\hline
\end{tabular}
\hfill{}
\end{table}
\end{center}
\end{theorem}

\begin{IEEEproof}
Due to the equivalency of the two problems, to solve (\ref{optorigin}), we consider (\ref{newopt}). Let us consider $i=2$. The case $i=1$ follows similarly. Due to \cite[Lemma 3]{Liu}, we need only consider the boundary points, i.e. $A$, $\mathbf{r}_2$ and $C$.

We now prove that point $\mathbf{r}_2=(r_1,r_2)$ cannot be the optimal solution if $w\neq w_2(\mathbf{r}_2)$. Suppose we draw a line across $\mathbf{r}_2$: $ar_1+br_2=1$ ($a,b\neq 0$). Evaluating $D_2$ (\ref{Di}) along this line by substituting $r_1=\frac{1-br_2}{a}$, we have
\begin{align*}
    D_2=w\frac{\tau_1}{R_1^*}+\frac{\tau_2(aR_1^*-w)}{aR_1^*r_2} + \frac{bw\tau_2}{aR_1^*}.
\end{align*}
If $a=\frac{w}{R_1^*}$, $D_2$ becomes a constant that is independent of $r_2$. This means that associated with any weight, there exists a line such that all feasible points on this line result in the same $D_2$. Now let us make this line tangent to $\mathcal{C}_1^G$: $a_tr_1+b_tr_2=1$. By Lemma \ref{tangent}, $w_2(\mathbf{r}_2)=a_tR_1^*\in(0,1)$. Note that $P_1<P$, hence $w_2(\mathbf{r}_2)<1$. For any weight $w\in(w_2(\mathbf{r}_2), 1]$, i.e. $a=\frac{w}{R_1^*}>a_t$, the associated line can be obtained by rotating the tangent line clockwise and hence it intersects $\mathcal{C}_{1,2}^G$. Referring to the Fig. \ref{BCCTRegion}(a), we have $D_2(\mathbf{r}_2)=D_2(\mathbf{r}'_2)$, where $\mathbf{r}'_2\in\mathcal{C}_1^G$ is some point on the line that is different from $\mathbf{r}_2$. Since $\mathbf{r}'_2$ is an inner point, it cannot be the optimal solution, so is $\mathbf{r}_2$. Similarly, if we consider any weight $w\in[0,w_2(\mathbf{r}_2))$, the associated line can be obtained by counter-clockwise rotating the tangent line and it intersects $\mathcal{C}_{1,2}^G$. Then we have $D_2(\mathbf{r}_2)= D_2(\mathbf{r}''_2)$ for some inner point $\mathbf{r}_2''$. Hence $\mathbf{r}_2$ cannot be the solution for any weight $w\in[0,w_2(\mathbf{r}_2))$.

Now let us consider $A$. For the same reason, we can show that $A$ cannot be the solution for any weight $w\in(w_2(A),1]$. The difference between $A$ and $\mathbf{r}_2$ is that since $A$ is the left-most boundary point, if we counter-clockwise rotate the tangent line at $A$, $A$ would still be the only feasible point on the line. Similarly, we can show that $C$ cannot be the solution for any weight $w\in[0,w_2(C))$.

To summarize, we have shown that $A$, $\mathbf{r}_2$ and $C$ may be the solution only if $w\in[0,w_2(A)]$, $w=w_2(\mathbf{r}_2)$ and $w\in[w_2(C),1]$ respectively. From the proof of Lemma \ref{tangent}, it's clear that $w_2(A)$, $w_2(\mathbf{r}_2)$ and $w_2(C)$ are all unique because $a$ is strictly (except the degenerate case $h_1= h_2$, for which Theorem \ref{miniBC} still holds) increasing w.r.t. $P_1$ . Hence $A$, $\mathbf{r}_2$ and $C$ are all associated with some disjoint sets of weights with which they may be the solution. Together with the fact that (\ref{newopt}) is solved at some boundary point for a given weight \cite[Lemma 3]{Liu}, we conclude that $A$, $\mathbf{r}_2$ and $C$ are indeed the solution for the weight(s) in their associated sets.
\end{IEEEproof}

\begin{corollary}
The minimum weighted sum completion time (\ref{optoverall}) for a weight $w$ is given by $d_s^*=\min\{D_1(\mathbf{r}_1^*), D_2(\mathbf{r}_2^*)\}$, where $w_i(\mathbf{r}_i^*)=w$.
\end{corollary}

Referring to Fig. \ref{BCCTRegion}(b), suppose the tangent line to curve $\bar{B}\bar{C}$ ($\bar{C}\bar{A}$) at point $\bar{C}$ has the slope of $s_1$ ($s_2$). Since all $\mathbf{d}\in\mathcal{D}_i^*$, $\mathbf{d}\neq \bar{C}$, are above the corresponding tangent line, $\bar{C}$ is the solution to the weighted sum completion time minimization problem over $\mathcal{D}_i^*$ with the weight $w=\frac{s_i}{s_i-1}$, $i=1,2$. We now use this geometric interpretation to prove that $\mathcal{D}^*$ is not convex.

\begin{proposition}
\label{BCnoconvex}
The completion time region $\mathcal{D}^*$ for Gaussian broadcast channel is not convex.
\end{proposition}
\begin{IEEEproof}
Referring to Fig. \ref{BCCTRegion}(a), suppose the tangent line across point $C$ is given by $ar_1+br_2=1$. Since $(R_1^*,R_2^*)$, where $R_i^*=\gamma(h_i^2P)$ for $i=1,2$, is above the line, we have $aR_1^*+bR_2^*>1$, or equivalently $w_1(C)< w_2(C)$. The tangent line to curve $\bar{B}\bar{C}$ or $\bar{C}\bar{A}$ at point $\bar{C}$ in Fig. \ref{BCCTRegion}(b) has the slope $s_i=\frac{w_i(C)}{w_i(C)-1}$, $i=1,2$ respectively. Therefore $s_1>s_2$ and $\mathcal{D}^*$ is not convex.
\end{IEEEproof}

\section{Completion Time Region for Gaussian Interference Channel}
Since this section mostly parallels Section III, for notation economy we may use the same notations introduced in Section III. Further, if not stated otherwise, the proofs of theorems are omitted and they follow their counterparts in Section III. Due to the fact that the capacity region of GIC is known only for certain ranges of channel parameters, we divide this section into three parts. In Section IV.A and IV.B, we consider the very strong and strong interference regimes respectively and establish the exact completion time region accordingly. In Section IV.C, we consider the weak and mixed interference regimes where achievable completion time region as well as outer-bound will be derived.

A general two-user GIC can be written equivalently in the following standard form:
\begin{align*}
    Y_1 &= X_1 + bX_2 + Z_1,\\
    Y_2 &= aX_1 + X_2 + Z_2,
\end{align*}
where $Z_i\sim\mathcal{N}(0,1)$, $i=1,2$, is the i.i.d. Gaussian noise process and inputs are subject to per symbol power constraints: $E[X_i^2]\leq P_i$. $a,b$ are some non-negative constants. Interference is said to be \textit{very strong} if $a\geq \sqrt{1+P_2}$ and $b\geq \sqrt{1+P_1}$. Interference is said to be \textit{strong} if $1\leq a< \sqrt{1+P_2}$ and $1\leq b< \sqrt{1+P_1}$. Interference is said to be \textit{weak} if $a< 1$ and $b<1$. Finally if one interference link is strong and the other is weak, then interference is said to be \textit{mixed}.

\subsection{Completion Time Region for GIC in the Very Strong Interference Regime}
\begin{theorem}
For a two-user GIC in the very strong interference regime, the $c$-constrained capacity region is the same as the standard capacity region and is given by $\{(R_1,R_2)|0\leq R_i\leq \gamma(P_i), i=1,2\}$.
\end{theorem}

The fact that interference is very strong enables each user to decode the interference by treating its own signal as noise and hence completely eliminate interference. As a result, the interference channel is decoupled into two point-to-point links and the fact that how many symbols one user's codebook spans does not really affect the coding scheme of the other.

\begin{theorem}
The completion time region of a two-user GIC in the very strong interference regime is $\mathcal{D}^*=\{(d_1,d_2)|d_i\geq \tau_i/\gamma(P_i),i=1,2\}$.
\end{theorem}

\subsection{Completion Time Region for GIC in the Strong Interference Regime}
\begin{theorem}
\label{GICstrongCapa}
The $c$-constrained capacity region of two-user GIC in the strong interference regime is the set of non-negative rate pairs $(R_1,R_2)$ satisfying:
\begin{enumerate}
\item{$c\leq 1$, $\left(R_1,\left[\tfrac{1}{c}R_2-(\tfrac{1}{c}-1)\gamma(P_2)\right]^+\right)\in\mathcal{C}_1^G$;}
\item{$c\geq 1$, $\left(\left[cR_1-(c-1)\gamma(P_1)\right]^+,R_2\right)\in\mathcal{C}_1^G$,}
\end{enumerate}
where $\mathcal{C}_1^G$ is the set of non-negative rate pairs satisfying
\begin{align*}
    r_1&\leq \gamma(P_1),\quad r_2\leq \gamma(P_2),\\
    r_1+r_2&\leq\min\{\gamma(P_1+b^2P_2),\gamma(a^2P_1+P_2)\}.
\end{align*}
\end{theorem}

In the strong interference regime, since the capacity achieving scheme requires each receiver to decode both the desired signal and the interference, the capacity region is equal to that of the compound MAC formed at the two receivers and resembles that of a GMAC except for the sum rate expression. The completion time region of GMAC was established in \cite{Liu} and can be directly transferred to the case of GIC in the strong interference regime by replacing the sum rate term $\gamma(P_1+P_2)$ by $r_{\textrm{s}}=\min\{\gamma(P_1+b^2P_2),\gamma(a^2P_1+P_2)\}$.

\begin{theorem}
The completion time region $\mathcal{D}^*$ of a two-user GIC in the strong interference regime is the set of pairs $(d_1,d_2)$ satisfying the following:
\begin{enumerate}
\item{For $\frac{\tau_2}{\tau_1}\leq \frac{r_{\textrm{s}}-\gamma(P_1)}{\gamma(P_1)}$}
\begin{align*}
    &\gamma(P_1)d_1\geq \tau_1,\ \gamma(P_2)d_2\geq \tau_2,\\
    &\gamma(P_1)d_1+[r_{\textrm{s}}-\gamma(P_1)]d_2\geq \tau_1+\tau_2.
\end{align*}
\item{For $\frac{r_{\textrm{s}}-\gamma(P_1)}{\gamma(P_1)}<\frac{\tau_2}{\tau_1}
    <\frac{\gamma(P_2)}{r_{\textrm{s}}-\gamma(P_2)}$}
\begin{align*}
    &\gamma(P_1)d_1\geq \tau_1,\ \gamma(P_2)d_2\geq \tau_2,\\
    &[r_{\textrm{s}}-\gamma(P_2)]d_1+\gamma(P_2)d_2\geq \tau_1+\tau_2,\\
    &\gamma(P_1)d_1+[r_{\textrm{s}}-\gamma(P_1)]d_2\geq \tau_1+\tau_2.
\end{align*}
\item{For $\frac{\tau_2}{\tau_1}\geq \frac{\gamma(P_2)}{r_{\textrm{s}}-\gamma(P_2)}$}
\begin{align*}
    &\gamma(P_1)d_1\geq \tau_1,\ \gamma(P_2)d_2\geq \tau_2,\\
    &[r_{\textrm{s}}-\gamma(P_2)]d_1+\gamma(P_2)d_2\geq \tau_1+\tau_2.
\end{align*}
\end{enumerate}
\end{theorem}

\subsection{Completion Time Region for GIC in the Weak and Mixed Interference Regimes}
Even though the exact capacity region of GIC in the weak and mixed interference regimes is still unknown, Etkin, Tse and Wang \cite{Etkin} derive a rate region that is at most one bit away from the outer-bound and hence establishes the capacity region to within one bit. Based on Etkin-Tse-Wang rate region, we derive the following achievable $c$-constrained rate region.

\begin{theorem}
\label{GICweakCapa}
The set of non-negative rate pairs $(R_1,R_2)$ satisfying the following constraints is an achievable $c$-constrained rate region of two-user GIC in the weak interference regime.
\begin{enumerate}
\item{$c\leq 1$, $\left(R_1,\left[\tfrac{1}{c}R_2-(\tfrac{1}{c}-1)\gamma(P_2)\right]^+\right)\in\mathcal{R}_{1,W}^G$;}
\item{$c\geq 1$, $\left(\left[cR_1-(c-1)\gamma(P_1)\right]^+,R_2\right)\in\mathcal{R}_{1,W}^G$.}
\end{enumerate}
Here $\mathcal{R}_{1,W}^G$ is the Etkin-Tse-Wang rate region for the weak interference regime and is given by \cite[Corollary.1]{Etkin}.
\end{theorem}

\begin{remark}
In the following, we focus on the weak interference regime. The obtained results are equally applicable to the mixed interference regime simply by replacing $\mathcal{R}_{1,W}^G$ with $\mathcal{R}_{1,M}^G$, the Etkin-Tse-Wang rate region for the mixed interference regime given by \cite[Corollary.2]{Etkin}.
\end{remark}

Generally, the rate region $\mathcal{R}_{1,W}^G$ is a polygon in the first quadrant with the dominant face consisting of line segments determined by the inequality constraints. To derive the completion time region, we follow the approach used in \cite{Liu} for MAC, where we first find the rate points that minimize the weighted sum completion time for $\mathcal{D}_1$ and $\mathcal{D}_2$ respectively and then their corresponding completion time pairs are the extreme points, by connecting which we can trace the boundary of $\mathcal{D}_1$ and $\mathcal{D}_2$ due to the convexity. Finally the achievable completion time region is given by the union of $\mathcal{D}_1$ and $\mathcal{D}_2$.

\begin{figure}[htb]
    \centering
    \includegraphics[width=90mm]{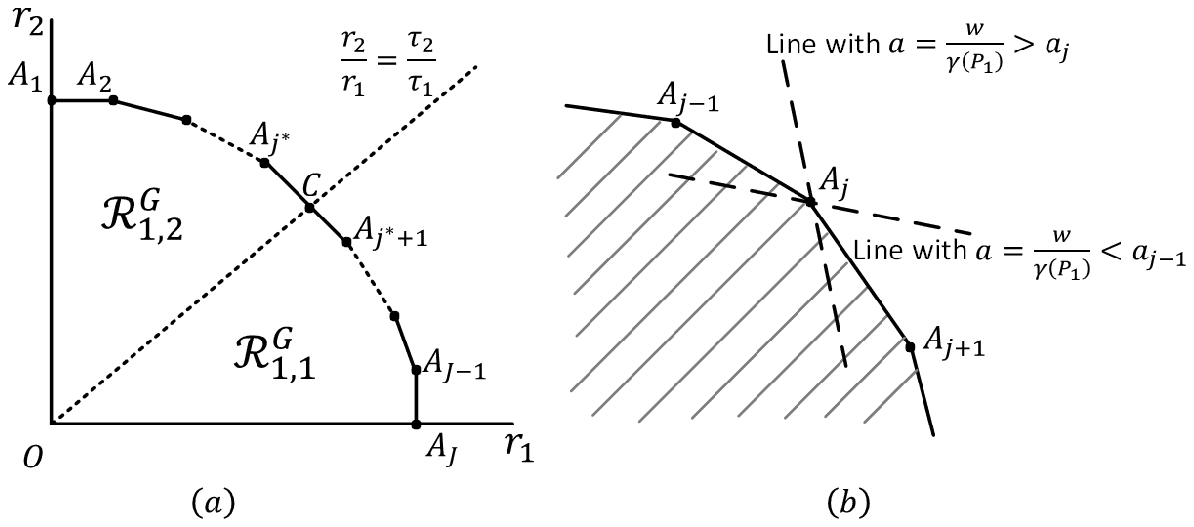}
    \caption{}
    \label{GeneralCTRegion}
\end{figure}

In \cite{Motahari}, the authors argue that all inequalities defining a Han-Kobayashi rate region for a given power splitting without time sharing, of which the Etkin-Tse-Wang rate region is a special case, are active. As observed in \cite{Xi}, under some channel conditions, certain inequalities are made redundant by the fact of rates being positive, which is neglected in \cite{Motahari}. Consequently, rate region $\mathcal{R}_{1,W}^G$ has a varying number of extreme points on the dominant face depending on the channel condition. We next consider a generic convex rate region $\mathcal{R}_1^G$, depicted in Fig. \ref{GeneralCTRegion}(a), whose boundaries are given by line segments $A_jA_{j+1}$, $j\in\{1,...,J-1\}$.

Let $C$ denote the point where line $r_2/r_1=\tau_2/\tau_1$ intersects the boundary of $\mathcal{R}_1^G$ and let
\begin{align*}
    j^*=\arg_{j\in\{1,...,J-1\}}\{C\in A_{j}A_{j+1}\}.
\end{align*}
Denote line segment $A_j A_{j+1}$ by $a_jr_1+b_jr_2=1$. Define $w^1_j\triangleq 1-\gamma(P_2)b_j$, $w^2_j \triangleq a_j\gamma(P_1)$. It can be shown that $a_1=0<a_2<...<a_{J-1}$ and $b_1>b_2>...>b_{J-1}=0$. Hence $w^1_1<w^1_2<...<w^1_{J-1}=1$ and $w^2_1=0<w^2_2<...<w^2_{J-1}$. Denote
\begin{align*}
    k_1^*&=\arg\min_{j\in\{j^*,...,J-1\}}\{w^1_j:w^1_j\geq 0\},\\
    k_2^*&=\arg\max_{j\in\{1,...,j^*\}}\{w^2_j:w^2_j\leq 1\}.
\end{align*}
Define two partitions for the unit interval $[0,1]$:
\begin{align*}
    \Pi_1&=[0, w^1_{k_1^*}],(w^1_{k_1^*},w^1_{k_1^*+1}],...,(w^1_{J-2},w^1_{J-1}],\\
    \Pi_2&=[w^2_1,w^2_2],...,(w^2_{k_2^*-1},w^2_{k_2^*}],(w^2_{k_2^*},1].
\end{align*}
Note that $w^1_{J-1}=1$ and $w^2_1=0$. Let $\Pi_i(j)$ denote the $j$th interval of partition $\Pi_i$.

\begin{lemma}
\label{Generaloptsol}
Consider the optimization problem (\ref{newopt}), where the feasible sets are replaced by $\mathcal{R}_{1,i}^G$, where
\begin{align*}
    \mathcal{R}_{1,1}^G&=\mathcal{R}_1^G\bigcap\{(r_1,r_2)|\tfrac{r_2}{r_1}\leq \tfrac{\tau_2}{\tau_1}\},\\
    \mathcal{R}_{1,2}^G&=\mathcal{R}_1^G\bigcap\{(r_1,r_2)|\tfrac{r_2}{r_1}\geq \tfrac{\tau_2}{\tau_1}\}.
\end{align*}

For any weight $w\in[0,1]$, let
\begin{align*}
    l_1^*&=\arg_{j\in\{1,...,J-k_1^*\}}\{w\in\Pi_1(j)\},\\
    l_2^*&=\arg_{j\in\{1,...,k_2^*\}}\{w\in\Pi_2(j)\}.
\end{align*}

Then $D_1$ is minimized at $A_{l_1^*+k_1^*-1}$ if $l_1^*\neq j^*+1-k_1^*$ or $C$ if $l_1^*= j^*+1-k_1^*$. Also $D_2$ is minimized at $A_{l_2^*+1}$ if $l_2^*\neq j^*$ or $C$ if $l_2^*= j^*$.
\end{lemma}

\begin{IEEEproof}
Let us consider $D_2$. The case for $D_1$ follows similarly. By \cite[Lemma.3]{Liu}, we need only consider extreme points on the dominant face of $\mathcal{R}_{1,2}^G$, i.e. $A_2$,...,$A_{j^*}$,$C$.
\begin{enumerate}
\item $k_2^*=j^*$, i.e. $w^2_{j^*}\leq 1$

For weight $w$ and $j\in\{2,...,j^*\}$, we can determine a line that passes $A_j$: $ar_1+br_2=1$ with $a=\frac{w}{\gamma(P_1)}$. From the proof of Theorem \ref{miniBC}, we know that all the feasible points on this line result in the same $D_2$. If $w\not\in [w^2_{j-1},w^2_{j}]$, the line intersects $\mathcal{R}_{1,2}^G$ and hence $A_j$ cannot be the minimizer. Equivalently, we say $A_j$, $j\in\{2,...,j^*\}$, may be the solution only if $w\in [w^2_{j-1},w^2_{j}]$. For the same reason, we can argue that $C$ may be the solution only if $w\geq w^2_{j^*}$. Similar to the proof of Theorem \ref{miniBC}, we can argue that $A_j$, $j\in\{2,...,j^*\}$, is indeed the solution if $w\in (w^2_{j-1},w^2_{j})$ and $C$ is the solution if $w> w^2_{j^*}$. When $w=w^2_{j}$, both $A_{j}$ and $A_{j+1}$ for $j\in\{2,...,j^*-1\}$ or $A_{j}$ and $C$ for $j=j^*$ are solutions and without loss of generality we pick the former.

\item $k_2^*<j^*$

Since the weight $w\leq 1$, we have $w\not\in[w^2_{j-1},w^2_{j}]$ for $k_2^*+1<j<j^*+1$. Hence we need only consider $A_j$, $j\in\{2,...,k_2^*\}$.
\end{enumerate}
\end{IEEEproof}

$\mathcal{R}_{1,W}^G$ is some convex rate region with boundaries given by piece-wise linear functions and there are at most six extreme points, i.e. $J\leq 6$. Hence we specialize Lemma \ref{Generaloptsol} to solve the optimization problem (\ref{newopt}) defined over $\mathcal{R}_{1,W}^G$. Denote the solutions for $D_i$ by $\{A_j\}_i$, $i=1,2$, $j\in\{1,...,J\}$, where the elements are sorted in ascending order of the subscript. Each element $A_j$ is associated with some set of weights with which $A_j$ minimizes $D_i$. Further let $\{\bar{A}_j\}_i$ denote the set of completion time points mapped from rate points $\{A_j\}_i$ using the mapping defined in (\ref{d1}) for $i=1$ and (\ref{d2}) for $i=2$ with $R_i''=\gamma(P_i)$. Let $\bar{C}$ be the point mapped from $C$, the intersection of $\mathcal{R}_{1,W}^G$ and the line $r_2/r_1=\tau_2/\tau_1$, using either (\ref{d1}) or (\ref{d2}) (the two mappings are equivalent for any rate point on the line $r_2/r_1=\tau_2/\tau_1$).

\textit{Construct $\mathcal{D}_i$ using $\{\bar{A}_j\}_i$ and $\bar{C}$:} $\mathcal{D}_i$ is a convex set of $\mathbf{d}\in \mathbb{R}_+^2$ whose boundaries consist of two rays and some line segments. The line segments are obtained by connecting the adjacent points in $\{\bar{C},\{\bar{A}_i\}_1\}$ for $\mathcal{D}_1$ and $\{\{\bar{A}_i\}_2,\bar{C}\}$ for $\mathcal{D}_2$. One of the two rays is the vertical (horizontal) ray emanating from $\bar{A}_{J-1}$ for $\mathcal{D}_1$ ($\bar{A}_2$ for $\mathcal{D}_2$). $\mathcal{D}_1$ and $\mathcal{D}_2$ share a common $45$ degree ray emanating from $\bar{C}$.

\begin{theorem}
$\mathcal{D}=\mathcal{D}_1\bigcup\mathcal{D}_2$ is an achievable completion time region for two-user GIC in the weak interference regime when Etkin-Tse-Wang coding scheme is used.
\end{theorem}
\begin{IEEEproof}
The proof is analogous to that of \cite[Theorem.3]{Liu}. Here we provide a sketch. We first argue that any point on the boundary of $\mathcal{D}_i$, $i=1,2$, is achievable when Etkin-Tse-Wang scheme is used. By the convexity of $\mathcal{D}_i$, any inner point is also achievable. Next we argue by contradiction that no rate point $\mathbf{r}\in\mathcal{R}_{1,W}^G$ can achieve $\mathbf{d}\not\in\mathcal{D}$. Suppose there exists $\mathbf{d}'\not\in\mathcal{D}$, then we could find a weight $w$ for which $\mathbf{d}'$ minimizes weighted sum completion time. But this contradicts with Lemma \ref{Generaloptsol}, which suggests that the solutions are the extreme points of $\mathcal{D}_i$.
\end{IEEEproof}

Replacing $\mathcal{R}_{1,W}^G$ by the capacity region outer-bound given by \cite[Theorem.3]{Etkin}, and following the above steps, we can obtain an outer-bound that includes the completion time region of GIC in the weak and mixed interference regimes.

\begin{figure}[htb]
    \centering
    \includegraphics[width=90mm]{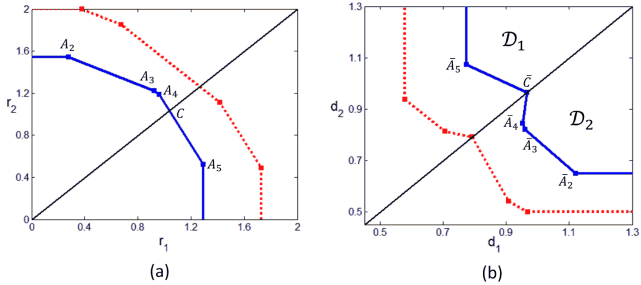}
    \caption{(a) The Etkin-Tse-Wang rate region (solid) and outer-bound (dashed); (b) achievable completion time region (solid) and outer-bound (dashed)}
    \label{GICWeakFig}
\end{figure}

\textit{Example:}
Let $P_1=10$, $P_2=15$, $a=0.8$, $b=0.6$ and $\tau_1=\tau_2=1$. The Etkin-Tse-Wang rate region and the outer-bound are depicted in Fig. \ref{GICWeakFig}(a). Let us compute $j^*=4$, $k_1^*=4$, $k_2^*=3$, $\Pi_1=[0, 0.36],(0.36, 1]$ and $\Pi_2=[0, 0.51],(0.51, 0.81],(0.81,1]$. By Lemma \ref{Generaloptsol}, the solutions set is $\{C,A_5\}$ for $D_1$ and $\{A_2,A_3,A_4\}$ for $D_2$. Therefore $\{\bar{C}, \bar{A}_5\}$, $\{\bar{A}_2,\bar{A}_3,\bar{A}_4,\bar{C}\}$ are the extreme points of $\mathcal{D}_1$ and $\mathcal{D}_2$ respectively. Fig. \ref{GICWeakFig}(b) plots the achievable completion time region and the outer-bound.

\section{Conclusions}
In this paper, we extend the information theoretic formulation of completion time, originally proposed in \cite{Liu} for multi-access channel, to broadcast channel and interference channel. This formulation allows us to define the so-called completion time region, which, analogous to capacity region, characterizes all possible trade-offs between users' completion times. The completion time region is established for Gaussian broadcast channel and is then proven to be non-convex by solving the weighted sum completion time minimization problem. For Gaussian interference channel, the exact completion time region is obtained for the very strong and strong interference regimes. For the weak and mixed interference regimes, an achievable completion time region based on the Etkin-Tse-Wang scheme and an outer-bound are obtained.

\appendices
\section{Proof of Theorem \ref{CCapaRegionBC}}
\label{ProofCCapaRegionBC}
\begin{IEEEproof}
We prove for $c\leq 1$, i.e. $n_1\leq n_2$. The case $c\geq 1$ follows similarly. The achievablility follows from Lemma \ref{raterelation}. Specifically for $R_2\leq (1-c)R_2^{\textrm{BC}}$, set $R_2'=0$ and $R_2''=\frac{1}{1-c}R_2$. For $R_2> (1-c)R_2^{\textrm{BC}}$, set $R_2'=\frac{1}{c}R_2-(\frac{1}{c}-1)R_2''$ and $R_2''=R_2^{\textrm{BC}}$.

The converse is as the following. Let $Q$ denote a uniformly distributed r.v. on $\{1,...,n_1\}$. For an arbitrarily small $\epsilon$,
\begin{align}
    n_1R_1&-n_1\epsilon\notag\\
    &\leq I(W_1;Y_1^{n_1})\label{bcr11}\\
    &\leq I(W_1;Y_1^{n_1}|W_2)\label{bcr13}\\
    &\leq n_1I(X_Q;Y_{1,Q}|U_Q,Q)\label{bcr14},
\end{align}
where (\ref{bcr11}) follows Fano's inequality, (\ref{bcr13}) is due to the fact that $W_1$ and $W_2$ are independent and conditioning reduces entropy. (\ref{bcr14}) follows standard steps for degraded broadcast channel \cite{Cover}. We proceed to bound $R_2$.
\begin{align}
    n_2R_2&-n_2\epsilon\notag\\
    &\leq I(W_2;Y_2^{n_2})\label{bcr21}\\
    &= I(W_2;Y_2^{n_1})+I(W_2;Y_{2,n_1+1}^{n_2}|Y_2^{n_1})\notag\\
    &\leq I(W_2;Y_2^{n_1})+I(W_2;Y_{2,n_1+1}^{n_2})\label{bcr22},
\end{align}
where (\ref{bcr21}) follows Fano's inequality, (\ref{bcr22}) is because $Y_{2,n_1+1}^{n_2}$ is independent of others conditioned on $W_2$ and conditioning reduces entropy. Following standard steps for degraded broadcast channel, we can show $I(W_2;Y_2^{n_1})\leq n_1 I(U_Q;Y_{2,Q}|Q)$. Also we have $I(W_2;Y_{2,n_1+1}^{n_2})\leq (n_2-n_1) R_2^{\textrm{BC}}$. Thus we have
\begin{align*}
    \tfrac{1}{c}R_2-(\tfrac{1}{c}-1)R_2^{\textrm{BC}}-\tfrac{1}{c}\epsilon\leq I(U_Q;Y_{2,Q}|Q).
\end{align*}
After redefining the r.v. $U\triangleq (U_Q,Q)$, the proof is complete.
\end{IEEEproof}

\section{Proof of Lemma \ref{tangent}}
\label{Prooftangent}
\begin{IEEEproof}
It's obvious $w_1< 1$ and $w_2> 0$, since $a,b>0$. For a boundary point $\mathbf{r}=(r_1,r_2)$, we have $r_1=\gamma(h_1^2P_1)$ and $r_2=\gamma(h_2^2P)-\gamma(h_2^2P_1)$. Since the slope of the tangent line at $\mathbf{r}$ is equal to $\frac{dr_2}{dr_1}=-\frac{a}{b}$, after some manipulation, we can show $\frac{a}{b}=g=\frac{1/h_1^2 + P_1 }{1/h_2^2 + P_1}$. Further because $ar_1+br_2=1$, we obtain $\frac{1}{b}=r_2+gr_1$ and $\frac{1}{a}=\frac{r_2}{g}+r_1$. Plugging in $r_1=\gamma(h_1^2P_1)$ and $r_2=\gamma(h_2^2P)-\gamma(h_2^2P_1)$, both $\frac{1}{a}$ and $\frac{1}{b}$ can be viewed as functions of $P_1$. After some manipulation, we can show that
\begin{align*}
    \tfrac{d}{dP_1}\left(\tfrac{1}{a}\right)&=-\tfrac{(h_1^2-h_2^2)h_1^2h_2^2}{(h_2^2+h_1^2h_2^2P_1)^2} \left[\gamma(h_2^2P)-\gamma(h_2^2P_1)\right]<0,\\
    \tfrac{d}{dP_1}\left(\tfrac{1}{b}\right)&=\tfrac{(h_1^2-h_2^2)h_1^2h_2^2}{(h_1^2+h_1^2h_2^2P_1)^2}\gamma(h_1^2P_1)>0.
\end{align*}
Since $\frac{1}{a}$ is monotonically decreasing on $P_1\in[0,P]$, we have $\frac{1}{a}|_{\textrm{min}}=\frac{1}{a}(P)=R_1^*$ and hence $w_2\leq 1$. Similarly $\frac{1}{b}$ is monotonically increasing on $P_1\in[0,P]$ and $\frac{1}{b}|_{\textrm{min}}=\frac{1}{b}(0)=R_2^*$. We have $w_1\geq 0$.
\end{IEEEproof}

\end{document}